\documentclass[11pt]{article}
\usepackage{graphicx}

\title{Light Hadron Masses in QCD with Valence Wilson Quarks at $\beta$=6.25 from a Parallel PC Cluster
\footnote{Supported by the
National Science Fund for Distinguished Young Scholars (19825117),
National Science Foundation, Guangdong Provincial Natural Science Foundation (990212), 
Ministry of Education, and Foundation of Zhongshan University Advanced Research Center.}
}
\vspace{2cm}
\author{Zhong-Hao MEI$^a$, 
Xiang-Qian LUO$^a$\thanks{Corresponding author. Email: stslxq@zsu.edu.cn. Phone: (8620) 84113928.}, 
Eric B. GREGORY$^{a,b}$\\
 {\small\sl $^a$ Department of Physics, Zhongshan University, Guangzhou 510275, China}\\
{\small\sl $^b$ Department of Physics, University of Arizona, Tucson, AZ 85721, USA}
}

\date{\today}

\begin{document}
\maketitle

\begin{abstract}
We present results for  $\pi$, $\rho$ and proton and $\Delta$  masses 
from our recently built Pentium cluster. 
The previous results for quenched Wilson fermions 
by MILC and GF11 collaborations are compared
at $\beta=6/g^2$=5.7 and 5.85 with the same parameters on 
the $8^3\times32$ lattice. 
New data are shown at $\beta$=6.25 on the
$12^3\times36$ and $16^3\times32$ lattices. 
Such a larger $\beta$ value is useful for extrapolating 
the lattice results to the continuum limit.
The smearing technique is systematically investigated and shown 
to greatly improve the spectrum data.
\end{abstract}



\setcounter{page}{0}
\newpage

\section{Introduction}
\label{sec1}

The computational physics group\cite{zsusite} at the Zhongshan University
is in a period of rapid development. The group's interests\cite{book,s1} 
cover such topics as lattice study of hadron spectroscopy\cite{Luo:1996,Luo:1997}, 
glueball decay and mixing, 
QCD at finite density\cite{Gregory:2000pm}, quantum instontons\cite{Jirari:1999bx} 
and quantum chaos\cite{Caron:2001zb} using the quantum action\cite{Jirari:1999ij}. 
Most of these topics can be studied through Monte Carlo simulation,
but can be quite costly in terms of computing power.  In order to do large scale
numerical investigations of these topics, we built a high performance parallel computer\cite{s2,s2b}
using PC components.  

QCD has been accepted to be the most successful gauge theory of strongly 
interacting particles.
Calculation of hadron spectroscopy remains to be an important task 
of non-perturbative studies of QCD using lattice methods. This paper is one 
of our first steps\cite{s1,s2} in the direction of studying hadron properties. Our reasons for 
performing simulations with Wilson valence quarks are twofold: 
First, we are interested in the analysis of hadronic spectroscopy for quenched Wilson 
fermions; 
Second, we are interested in exploring the performance of our cluster 
in actual lattice QCD simulations. We decided to do a spectrum calculation on
the $8^3\times32$ lattice at $\beta=5.7$ and 5.85, and on the $12^3\times 36$ and $16^3\times32$  
lattices at $\beta=6.25$. The results at smaller $\beta$ values
are used to compare with those in the literature, while those 
at a larger $\beta$ value, corresponding to smaller lattice spacing $a$, 
are useful for extrapolating the lattice results to the continuum limit.
The smearing technique is employed to improve the spectrum data.
We hope our data will be an important addition to the lattice 
study of QCD spectrum.

   The remainder of this paper is organized as follows.  
 In Sect. \ref{sec2}, the 
basic ideas of lattice QCD with Wilson quarks are given.
In Sect. \ref{sec3}, 
we describe the simulation parameters  
as well as some basic features of our cluster. 
In Sect. \ref{sec4}, we present new
calculation of some light hadron masses.
Finally, conclusions and 
outlook are given in Sect. \ref{sec5}.

\section{Basic ideas of lattice QCD}
\label{sec2} 

Our starting point is the Wilson action \cite{wilson}
\begin{eqnarray}
S=-{\beta \over 6} \sum_p {\rm Tr} (U_{p} +U_{p}^{\dagger}-2)+\sum_{x,y} \bar{\psi}(x) M_{xy}\psi(y),
\label{action}
\end{eqnarray}
where $\beta=6/g^2$, and  $U_p$ is the ordered product 
of gauge link variables $U_{\mu}(x)=e^{ig\int_{xa}^{xa+ \hat{\mu}a} dx' A_{\mu}(x')}$ around an elementary 
plaquette. $a$ is the lattice spacing, $\hat{\mu}$ is the unit vector and  $M$ is the fermionic matrix:
\begin{eqnarray}
    M_{xy}=\delta_{xy}-
\kappa \sum_{\mu=1}^{4} \left[ (1-\gamma_{\mu})U_{\mu}(x)\delta_{x,y- \hat{\mu}}
+
(1+\gamma_{\mu})U_{\mu}^{\dagger}(x- \hat{\mu})\delta_{x,y+ \hat{\mu}} \right].
\end{eqnarray}
The fermion field $\psi$ on the lattice is related to the continuum one $\psi^{cont}$ 
by $\psi=\psi^{cont} \sqrt{a^3/(2\kappa)}$ with $\kappa=1/(2ma+8)$. 
Then all the physical quantities are calculable through 
Monte Carlo (MC) simulations with importance sampling.
Fermion fields must be integrated out before the simulations, leading to
the vacuum expectation value for an operator $F$
\begin{eqnarray}
\langle F \rangle = {\int [dU] \bar{F}([U]) e^{-S_{g}([U])} 
\left( \det M \right)^{N_f} 
\over \int [dU] e^{-S_g([U])}\left( \det M \right)^{N_f}}
\approx {1 \over N_{\rm config}} \sum_{\mathcal C}  \bar{F}[{\mathcal C}]. 
\end{eqnarray}
Here $N_f$ is the number of flavors, and $\bar{F}$ is the corresponding 
operator after Wick contraction of the fermion fields
and the summation is over the gluonic configurations, ${\mathcal C}$, drawn 
from the Boltzmann distribution.  In quenched approximation, $\det (M)=1$.

The correlation functions of a hadron is:
\begin{eqnarray}
   C_{h}(t)=\sum_{\vec{x}}
\langle O_h^{\dagger}(\vec{x},t)O_h(\vec{0},0)\rangle,
\end{eqnarray}
where $O_h(\stackrel{\rightarrow}{x},t)$  
is a hadron operator. 
For sufficiently large values of $t$ and the lattice time period $T$,
the correlation function is expected to approach the asymptotic form:
\begin{eqnarray}
   C_h(t)\rightarrow Z_{h}[\exp(-m_{h}at)+\exp(m_{h}at-m_{h}aT)].
\label{fit}
\end{eqnarray} 
Fitting the large $t$ behavior of the correlation function according to Eq. (\ref{fit}), 
the hadron mass $m_h$ is  obtained.

\section{Simulations}
\label{sec3}

The quenched gauge field configurations 
and quark propagators were obtained using
the recently built ZSU cluster\cite{s2,s2b}. 
The full machine has 10 nodes, and each node consists of dual Pentium CPUs.
The total internal memory is 1.28GB 
and the sustained speed is around 2 Gflops. 
An upgrade is planned.
The cluster runs on the Linux operating system. 
To operate the cluster as a parallel computer, the
programmer must design the algorithm so that it appropriately divides the
task among the individual processors. We use MPI (Message Passing Interface), 
one of the most popular message passing standards.

We updated the pure SU(3) gauge fields
with Cabibbo-Marinari quasi-heat bath algorithm\cite{Cabibbo:1982zn}, 
each iteration followed by 4 over-relaxation sweeps.
The simulation parameters are listed in Tab. \ref{parameters}.
The distance between two nearest stored configurations is 100.
The auto-correlation time was computed 
to make sure that these configurations  are independent.

The $u$ quark and $d$ quark are assumed to be degenerate.
The quark propagators 
are calculated in the Coulomb gauge 
using the independent configurations mentioned above, and
conjugate gradient
for inversion of the Dirac matrix with preconditioning via
ILU decomposition by checkerboards\cite{s9}.

At $\beta=6.25$,  we computed the $\pi$, $\rho$ and proton and $\Delta$ 
masses on the $12^3\times36$ and $16^3\times 32$ lattices 
at four values of hopping 
parameter: $\kappa=0.1480$, 0.1486,  0.1492,  and 0.1498  
with point and smeared\cite{s3,s7} sources.
In order to compare our results with those by MILC and GF11,
the quenched simulations were also performed at 
$\beta=5.7 $ and $ \beta=5.85$  on the  $8^3\times32$ lattice.
We repeated the  quenched simulations of the
MILC and GF11 collaborations, using the same set of $\kappa$ 
values \cite{s4,s3,s7} but on a $8^{3}\times 32$ lattice.
Detailed data will be presented elsewhere \cite{s2}.

\section{Light hadron masses}
\label{sec4}

To extract masses from the hadron propagators, we must average the 
correlation function of the hadron over the ensemble of gauge configurations, 
and use a fitting routine to evaluate the hadron masses $m_h$.
We determined hadron masses by fitting our data under the assumption
that there is a single particle in each channel \cite{s2}.

Point source means a delta function, and smeared
source means a spread-out distribution 
(an approximation to the actual wave-function of the quantum state).
For example, the
simplest operator for a meson is 
just $O_h(\stackrel{\rightarrow}{x})={\bar q}(\vec{x}) q(\vec{x})$, 
i.e. the product of quark and anti-quark
fields at a single point.
A disadvantage of this point source,  
is that this operator creates not only
the lightest meson, but all possible excited states
too.  To write down 
an operator which creates more of the single state, 
one must ``smear" the operator out, e.g.
$\sum_{\vec{y}} ~ {\bar q} (\vec{x}) f(\vec{x}-\vec{y}) q(\vec{y})$ 
where $f(\vec{x})$ is some smooth function. Here we choose 
\begin{eqnarray}
f(\vec{x}) = N\rm{exp}(-|\vec{x}|^2/r_0^2),
\end{eqnarray}
with $N$ a normalization factor.  
The size of the smeared operator
should generally be comparable to the size of the hadron created.
There is no automatic procedure \cite{s3}
for tuning the smearing parameter $r_0$.  One simply has to
experiment with a couple of choices.

\begin{figure}[hb]
\begin{center}
\rotatebox{270}{\includegraphics[width=8cm]{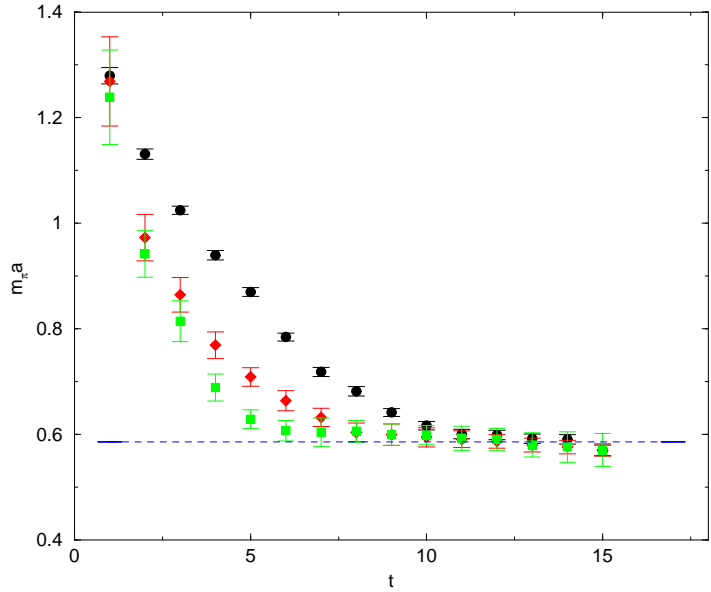}}
\end{center}
\caption{\label{pion_6.25_36_48}  Pion effective mass fits to 
the correlation function at  $\beta=6.25$ and $\kappa=0.1480$
and on the $16^3\times32$ lattice.
Data for the point source, smearing source for $r_0=8$, and $r_0=26$ are labelled 
by circles, diamonds, and squares respectively (from top to bottom).}
\end{figure}

\begin{figure}[hb]
\begin{center}
\rotatebox{270}{\includegraphics[width=8cm]{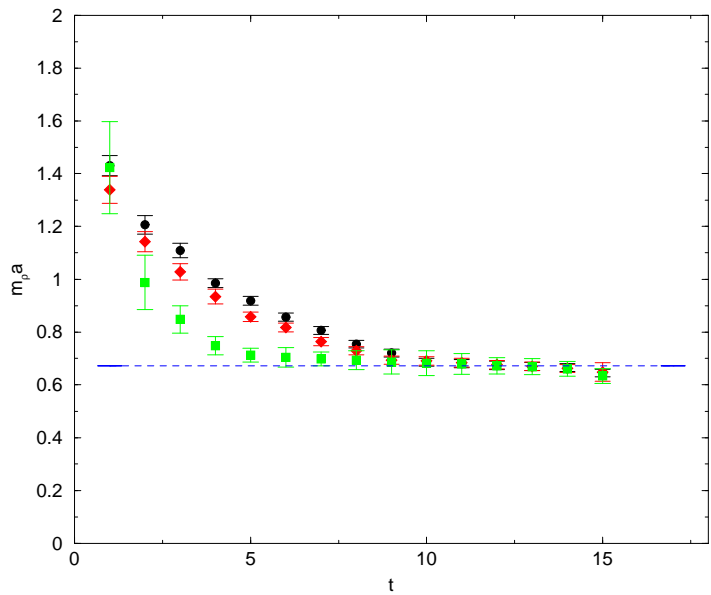}}
\end{center}
\caption{\label{rho_6.25_36_48} The meaning of the symbols are the same as Fig. \ref{pion_6.25_36_48}, but the rho meson.
}
\end{figure}

The effective meson mass for $\pi$ and $\rho$ as a function of time $t$ at
$\beta=6.25$ and on the $16^3\times32$ lattice is depicted
in Figs. \ref{pion_6.25_36_48} and
\ref{rho_6.25_36_48} respectively. 
As one sees, the plateau from which one can estimate the mass, 
is very narrow for point source, due to the reason mentioned above.
When the smearing source is used, 
the width of the plateau changes with the smearing parameter $r_0$.
We tried many values of $r_0$ and found that when $r_0>16$, the effective mass is almost independent on $r_0$ and
the optimal value for $r_0$ is 26, where we observe the widest plateau.

\begin{figure}[hb]
\begin{center}
\rotatebox{270}{\includegraphics[width=8cm]{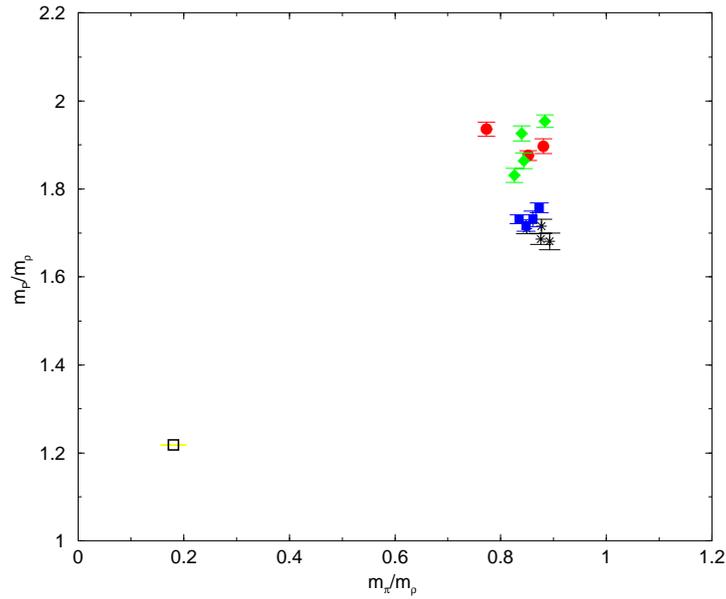}}
\end{center}
\caption{\label{edinburgh_5.7_5.85_6.25} Edinburgh plot. 
Data are: simulations at $\beta=5.7$ (star), $\beta=5.85$ (filled circle) on 
the $8^3\times32$ lattice,  $\beta=6.25$ on the $12^3\times36$ lattice (diamond)
and $\beta=6.25$ on the $16^3\times 32$ lattice (filled square).
The open square shows the real-world point.}
\end{figure}

The $\pi$, $\rho$, proton and $\Delta$ masses \cite{s2} at $\beta=5.7$ and on $8^3\times32$ are consistent with 
the previous GF11 data \cite{s7}  and those for the mesons at $\beta=5.85$ and on $8^3\times32$
are comparable with  those in \cite{s7} on the
$16^3\times32$ lattice. This means the finite size effects are very small.
At $\beta=5.85$, we also see that our results for the proton and $\Delta$ masses\cite{s2} 
are bigger than
the MILC data and consistent with the finite size behavior analysis 
in \cite{s3}. 
More detailed results will be given in \cite{s2}.
In Fig. \ref{edinburgh_5.7_5.85_6.25}, we show the Edinburgh plot, 
($m_N/m_p$) vs. ($m_\pi/m_\rho$) mass 
ratios.

\begin{figure}[hb]
\begin{center}
\rotatebox{270}{\includegraphics[width=8cm]{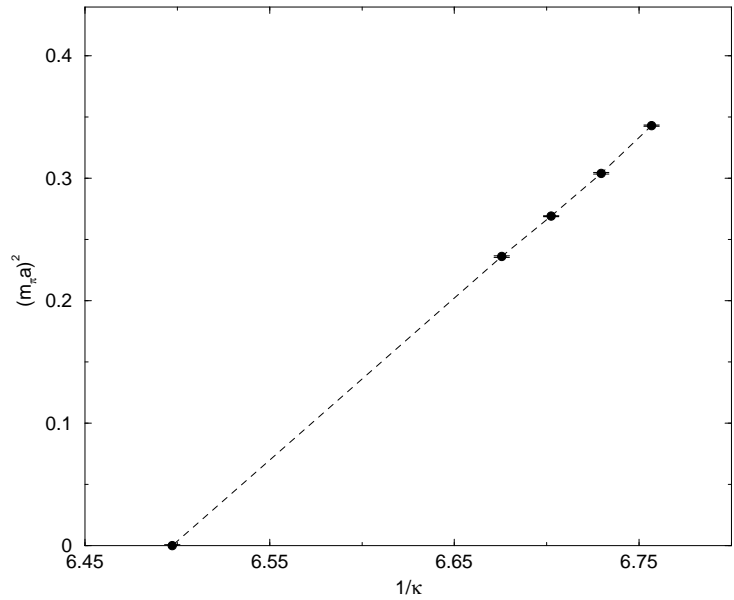}}
\end{center}
\caption{\label{pion_kappa_6.25}  Pion mass squared as a function of 
 $1/\kappa$ for  $\beta=6.25$ on the $16^3\times 32$ lattice.  In this and the 
following figures, the error bars reflect the sum of statistical and 
systematic errors.}
\end{figure}

\begin{figure}[hb]
\begin{center}
\rotatebox{270}{\includegraphics[width=8cm]{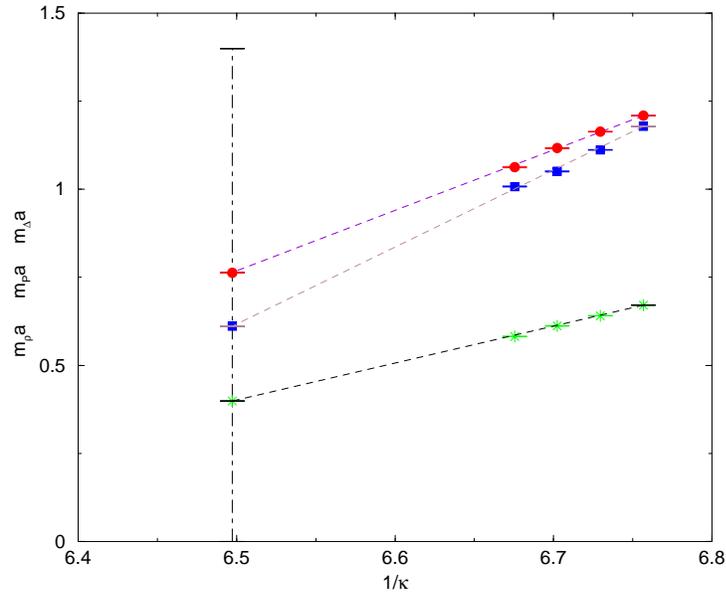}}
\end{center}
\caption{\label{rho_kappa_6.25} Rho, proton and Delta masses as a function of 
 $1/\kappa$ for  $\beta=6.25$ on the $16^3\times 32$ lattice. 
Results in the chiral limit (on the vertical dot-dashed line) are also shown.}
\end{figure}

In Figs.  \ref{pion_kappa_6.25}, and \ref{rho_kappa_6.25}, 
we show respectively $(m_{\pi}a)^{2}$, $m_{\rho}a$, $m_{p}a$ 
and $ m_{\Delta}a$ 
as a function of $1/\kappa$ at $\beta=6.25$ and on the $16^3\times 32$ lattice.
Assuming that $(m_{\pi}a)^{2}$ is linear in $ 1/\kappa$, we can compute the 
critical coupling $\kappa_{c}$ at which the pion becomes massless. We 
extrapolate the data using:
\begin{equation}
 \label{e3}
(m_{\pi}a)^{2}=A \left({\frac{1}{\kappa}}-{\frac{1}{\kappa_{c}}}\right).
\end{equation}
The results for $\kappa_c$, and $m_{\rho}a$, $m_{P}a$ 
and $m_{\Delta}a$ at $\kappa_c$ on the $12^3\times 36$ and $16^3\times 32$ are given in Tab.\ref{Tab2},
where the experimental value for the $\rho$ mass has been used as an input.
The finite size effects are consistent with \cite{s3}. Also, the mass ratios $m_P/m_{\rho}$
and $m_{\Delta}/m_{\rho}$ seem to be closer the experimental values, 1.222 and 1.604 on the larger lattice.  
To compare $m_P$ and $m_{\Delta}$ with experiment, we need to do simulation on a larger lattice volume 
and carefully study the lattice spacing errors. 
In this aspect, it might be more efficient to use the 
improved action and some progress\cite{Wu,Liu} has been reported by  
some lattice groups in China.

\section{Conclusions and Outlook}
\label{sec5}

  In this paper, we have presented new data on hadron masses in QCD with 
Wilson valence quarks at $\beta=6.25$ on the $12^3 \times 36$ and $16^3\times 32$ lattices. 
Our results at $\beta=5.7$ are consistent and those at $\beta=5.85$ 
are comparable with the results in the literature.
We have also made a more careful and systematic study of the smearing method.
Such large scale simulations had usually required supercomputing resources \cite{s3,s7},
but now they were all done on our recently built cluster for parallel
computer.
  In the task of lattice QCD simulations, we are confident that ZSU's Pentium 
cluster can provide a very flexible and extremely economical computing 
solution, which fits the demands and budget of a developing lattice field 
theory group. We are going to use the machine to produce more
useful results of non-perturbative physics.

\bigskip

\noindent
{\bf Acknowledgements}

This work was in part based on the MILC collaboration's public 
lattice gauge theory code. (See reference \cite{milc}.)  We are grateful to 
C. DeTar, C. McNeile, and D. Toussaint, for helpful discussions.

\begin{table}
\begin{center}
\begin{tabular}{|c|c|c|c|}\hline
  volume          &  $\beta$ & warmup & stored configs.  \\ 
\hline
   $8^3\times 32$ &    5.7   &    200  & 200       \\ 
\hline
   $8^3\times 32$ &    5.85  &    200  & 200        \\ 
\hline
  $12^3\times 36$ &    6.25  &    200  & 200         \\ 
\hline  
$16^3\times 32$   &    6.25  &    600  & 600         \\ 
\hline               
\end{tabular}
\end{center}
\caption{\label{parameters} Simulation parameters}
\end{table}

\bigskip

\begin{table}
\begin{center}
\begin{tabular}{|c|c|c|}\hline
Lattice Volume          & $12^3\times 36$ & $16^3\times 32$ \\ \hline
$\kappa_c$              & 0.1531(7)           & 0.1539(11)          \\ \hline
$a(\kappa_c)$           & 0.505(9) GeV$^{-1}$ & 0.520(5) GeV$^{-1}$ \\ \hline
$m_{\pi}a(\kappa_c)$    & 0.0                 & 0.0                 \\ \hline
$m_{\rho}a(\kappa_c)$   & 0.388(6)            & 0.399(8)            \\ \hline
$m_{P}a(\kappa_c)$      & 0.621(8)            & 0.611(5)            \\ \hline
$m_{\Delta}a(\kappa_c)$ & 0.770(3)            & 0.763(8)            \\ \hline
$m_P/m_{\rho}$          &  1.598(4)           & 1.529(8)            \\ \hline
$m_{\Delta}/m_{\rho}$   &  1.981(9)           & 1.908(9)            \\ \hline
\end{tabular}
\caption{\label{Tab2} Results of fitted $\kappa$, lattice spacing, and light hadron masses in the
chiral limit at $\beta$=6.25.}
\end{center}
\end{table}

\end{document}